\documentclass[twocolumn,showpacs,amsmath,amssymb,showkeys,floatfix,prl]{revtex4}
\usepackage{amsfonts,amsmath}
\usepackage{graphicx}
\usepackage{dcolumn}
\usepackage{bm}

\begin{document}

\title{Impact of approximate oscillation probabilities in the analysis of three neutrino experiments}

\author{B. K. Cogswell$^1$, D. C. Latimer$^2$ and D. J. Ernst$^1$}

\affiliation{$^1$Department of Physics and Astronomy, Vanderbilt University,
Nashville, Tennessee 37235}

\affiliation{$^2$Department of Physics, University of Puget Sound, Tacoma, Washington 93416}

\begin{abstract}
As neutrino oscillation data becomes ever more precise, the use of approximate formulae for the oscillation
probabilities ${\mathcal P}_{\alpha\beta}$ must be examined to ensure that the approximation is adequate. Here, the
oscillation probability ${\mathcal P}_{ee}$ is investigated in the context of the Daya Bay experiment; the oscillation
probability ${\mathcal P}_{\mu\mu}$ is investigated in terms of the T2K disappearance experiment; and the probability ${\mathcal P}_{\mu
e}$ is investigated in terms of the T2K appearance experiment. Daya Bay requires ${\mathcal P}_{ee}$ in vacuum and
thus the simple analytic formula negates the need for an approximate formula. However, improved data from T2K will soon
become sensitive to the hierarchy, and thus require a more careful treatment of that aspect. For the other cases, we choose an
expansion by Akhmedov {\em et al.~}which systematically includes all terms through second order in $\sin\theta_{13}$ and in
$\alpha =: \Delta_{21}/\Delta_{31}$ ($\Delta_{jk} =: m^2_j - m^2_k$). For the T2K
disappearance experiment the approximation is quite accurate. However, for the T2K appearance experiment the
approximate formula is not precise enough for addressing such questions as hierarchy or the existence of CP violation in the lepton sector. We suggest
the use of numerical calculations of the oscillation probabilities, which are stable, accurate, and efficient and
eliminate the possibility that differences in analyses are emanating from different approximations. 
\end{abstract}



\maketitle

\section{Introduction}

The oscillation of three neutrino flavors can be parameterized in terms of six real independent parameters:  three mixing angles $\theta_{jk}$, two mass-squared differences $\Delta_{jk}:= m_j^2 - m_k^2$, and the Dirac CP phase 
$\delta_{CP}$.   The goal of neutrino oscillation experiments is to measure these parameters with high precision.  Broadly, these experiments fall into one of two categories: single-parameter constraint or multi-parameter 
sensitivity. Through shrewd experimental design or wise choice of baseline or neutrino energy, an experiment
can be predominantly sensitive to only one of these parameters.  That is, the experiment can cleanly extract the value of a single parameter with little sensitivity to the precise values of the other parameters.  As examples, 
the Daya Bay \cite{An:2013zwz} experiment provides a clean measurement of the mixing angle $\theta_{13}$, and long-baseline muon disappearance experiments, like MINOS  \cite{Adamson:2014vgd}  and T2K \cite{Abe:2013fuq}, are 
able to determine the mass-squared difference $\Delta_{32}$ with minimal knowledge of other parameters.

On the other hand, experiments designed to measure other oscillation properties, like the ordering of the mass
eigenstates or the value of the CP phase, are particularly sensitive to the values of the oscillation parameters
determined by other experiments.
As an example, measurements of electron neutrino appearance in a muon neutrino beam, such as with T2K and NO$\nu$A
\cite{Jediny:2014lda}, can be used to ascertain the neutrino mass hierarchy or the existence of CP violation, but 
the extraction of these features relies heavily upon the mixing parameters measured by other experiments. The interdependent sensitivity of these extracted parameters on other parameters requires a careful understanding of 
how commonly used approximate neutrino oscillation formulae may affect the outcome of an analysis.

Herein, we examine the adequacy of utilizing approximate formulae for the oscillation probabilities in the analysis of
neutrino oscillation experimental data. For approximate formulae, we adopt the highly cited perturbative expansion proposed in 
Ref.~\cite{Akhmedov:2004ny}, which systematically incorporates all terms of second-order in the small quantities $\sin\theta_{13}$ and the ratio of the mass-squared differences $\alpha:= \Delta_{21}/\Delta_{31}$.   To  numerically compute the exact oscillation probabilities in matter, we use the 
method of Ohlsson and Snellman
\cite{Ohlsson:1999xb}.
We consider three representative experiments:  Daya Bay
\cite{An:2013zwz}, the T2K disappearance experiment \cite{Abe:2013fuq}, and the T2K appearance experiment
\cite{Abe:2013hdq}.  When evaluating the utility of an approximation, of primary importance is the accuracy of the extracted oscillation parameters, not solely the accuracy of the oscillation probability. Given this, we extend our investigation beyond the usual assessment of deviations between probabilities to deviations between statistical outcomes based on differing probability formulae.

\section{ The Daya Bay Experiment and $\theta_{13}$}

In this section, we examine the mixing angle $\theta_{13}$ in the context of the Daya Bay experiment.   The exact
formula for the vacuum electron neutrino survival probability is
\begin{eqnarray} 
{\mathcal P}_{ee} &=& 1 \nonumber\\
                  &-& \sin^2 2\theta_{13}\,(c^2_{12}\,\sin^2\phi_{31}+s^2_{12}\,\sin^2\phi_{32})\nonumber\\
                  &-&c^4_{13}\,\sin^2 2\theta_{12} \,\sin^2\phi_{21}\,\,,
\label{exactp}
\end{eqnarray}
with $s_{jk}:=\sin\theta_{jk}$, $c_{jk}:=\cos \theta_{jk}$, and $\phi_{jk}:=1.267\,\Delta_{jk} L/E$, where the baseline $L$
is in meters, the neutrino energy $E$ is in  MeV, and the mass-squared differences $\Delta_{jk}$ are in eV$^2$. 
Although the exact vacuum oscillation formula is simple and easy to use, we learn some things by investigating its
validity.
 
\begin{figure}
\includegraphics*[width=3in]{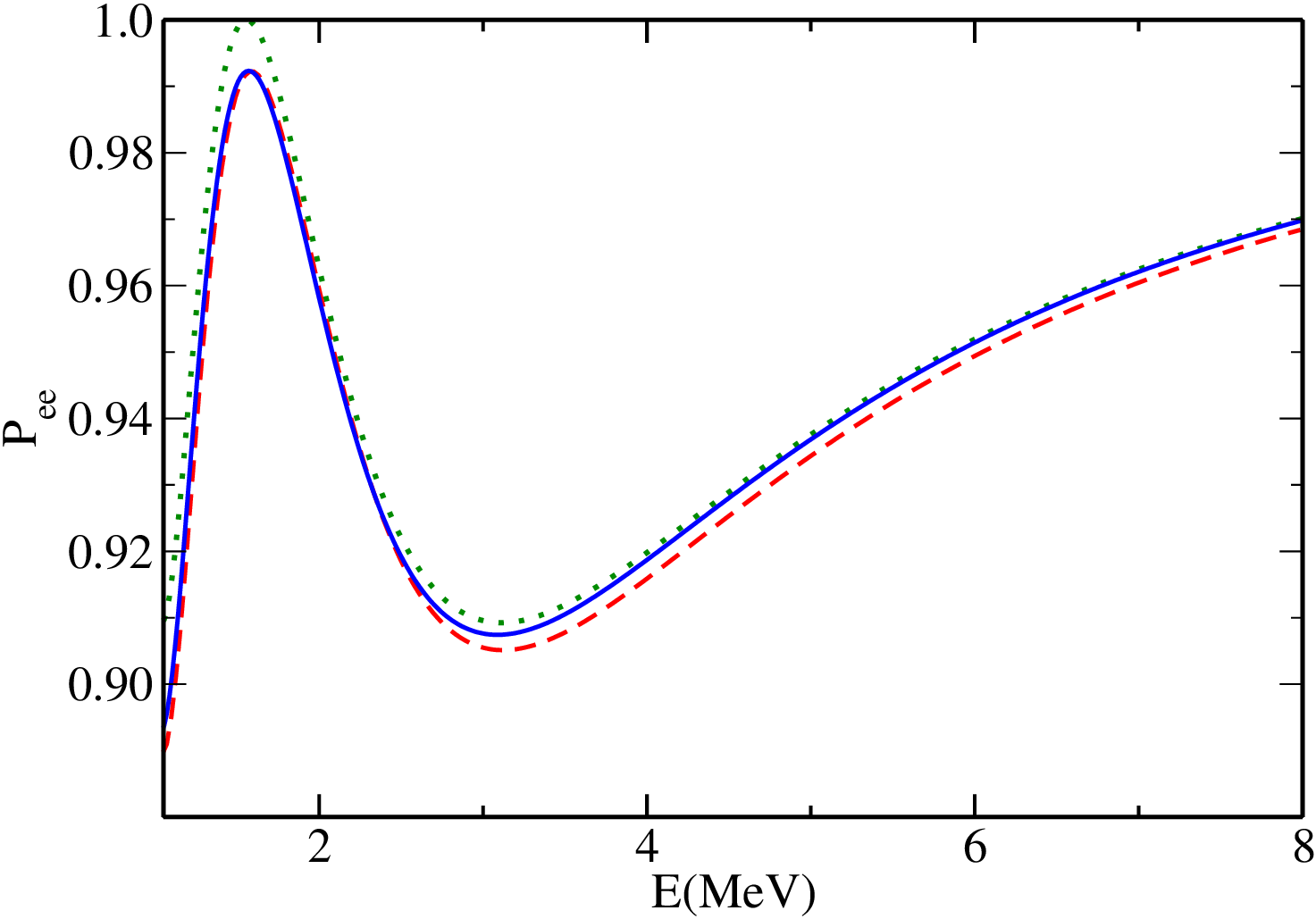}
\caption{The oscillation probability ${\mathcal P}_{ee}$ as a function of $E$ for a baseline of $L=1626$ m. The solid (blue) line represents multiple indistinguishable calculations:  the exact vacuum oscillation probability 
for either hierarchy,  the exact oscillation probability through matter of density 2.65 g/cm$^3$, and the
$\Delta_{ee}$ approach including the $\phi_{21}$ term from \protect Eq.~\ref{exactp}. The dotted (green) curve represents the exact probability minus the last term in 
Eq.~(\ref{exactp}). The dashed (red) curve is the result of the approximation of Ref.~\cite{Akhmedov:2004ny}.  }
\label{fig1}
\end{figure}

The Daya Bay experiment measures nuclear reactor electron antineutrinos over a baseline of $L=1626$ m, the flux-averaged distance from the reactors to the far detector.  It is the dominant experiment in the determination of 
$\theta_{13}$.   In Fig.~\ref{fig1}, we use exact and approximate oscillation formulae to plot  the survival probability ${\mathcal P}_{ee}$  for an energy range relevant to the Daya Bay experiment. 

We can check the accuracy of our computer code by comparing its results with the exact oscillation formula
Eq.~(\ref{exactp}) for vacuum oscillations. The numerical calculation reproduces the analytic formula to the accuracy of
the computer, for us, fifteen decimal places. The first question is whether the effects of the interactions of the neutrino with the Earth's matter may be neglected. Utilizing a typical matter density of 2.65 g/cm$^3$, we 
find that the
percent change in ${\mathcal P}_{ee}$ caused by the interaction with matter has two peaks, 0.003\% at 1.28 MeV and
0.007\% at 2.41 MeV. Matter effects are thus negligible. The second question is whether
hierarchy can be neglected. Hierarchy refers to the two cases of the mass ordering: normal hierarchy is the situation
when the third mass state has a larger mass than the two other states and inverse hierarchy refers to the case when the
third mass state has a mass smaller than the masses of the two other states. We move from the normal to  the inverse hierarchy via the map
$\Delta_{31}\mapsto -\Delta_{31}+ \Delta_{21}$.
Changing from normal to inverse hierarchy changes ${\mathcal
P}_{ee}$ by 0.32\% at 2.00 MeV and by 0.12\% at 4.73 MeV. We
here adopt a convention that if the calculation is better than one percent accurate, it is acceptable. Thus neglecting matter
and hierarchy effects is acceptable when considering the oscillation probability.  The solid (blue) curve in Fig.~\ref{fig1}
represents the vacuum results, the inclusion of matter effects, and the results for either
hierarchy. 

The dotted (green) curve is the result of omitting the small $\phi_{21}$
term in Eq.~(\ref{exactp}), assuming normal hierarchy and vacuum oscillations. Historically, this term would be referred to as a subdominant effect. This produces maxima in
the percent error of 0.05\% at 6.09 MeV and 0.8\% at 1.80 MeV, just above the peak.  At the peak, the muon interaction
rate has become so small that this term has little effect. Although it could be ignored, there is no reason to do so.

The approximate formula for the vacuum electron neutrino survival probability from Ref.~\cite{Akhmedov:2004ny}  is
\begin{equation}
{\mathcal P}_{ee} = 1-4 s^2_{13} \sin^2\phi_{31} - \alpha^2 \phi_{31}^2 \sin^2 2\theta_{12} ,
\label{appee}
\end{equation}
valid to second order in both $\sin\theta_{13}$ and  $\alpha$.
This produces the dashed (red) curve in Fig.~\ref{fig1}. The peak percent difference for this approximation from the exact
probability is 0.1\% at an energy
of 1.80 MeV (near the peak) and 0.3\% at 3.76 MeV. 

We also examine an approximation used by the Daya Bay collaboration. In this
approximation, the oscillations driven by $\Delta_{31}$ and $\Delta_{32}$ are replaced  by a single oscillation driven by the  mass-squared difference $\Delta_{ee}$ \cite{Minakata:2006gq}. This effective mass-squared difference 
$\Delta_{ee}$ is determined by requiring that the exact and approximate oscillation minima are equal. The result is
\begin{equation}
\Delta_{ee} = c^2_{12}\,\Delta_{31}+s^2_{12}\,\Delta_{32}\,\,.
\end{equation}
The $\sin^2 \phi_{3j}$ terms in Eq.~(\ref{exactp}) are averaged by a weighting of $c^2_{12}$ and $s^2_{12}$.
The approximation, Eq.~(\ref{appee}), simply transfers this weighted averaging to the mass-squared differences, $\Delta_{3j}$. The difference from the exact result is everywhere less than 0.02\%, a very good approximation, and this 
result is also depicted  by the blue curve in Fig.~\ref{fig1}.

It is not the accuracy of the oscillation probability that determines whether an approximation is adequate; rather, the ability to extract accurate oscillation parameters from data is the bottom line in determining an 
approximation's utility. 
Thus, in our analysis of the Daya Bay experiment, we compare the extracted values for $\theta_{13}$ that result from the various exact and approximate oscillation formulae. In this analysis, we fix  the mixing parameters to their measured values \cite{Beringer:1900zz}: $\theta_{12}=0.557$, $\Delta_{21}=7.5\times
10^{-5}$ eV$^2$; and we consider maximal mixing for the atmospheric angle:  $\theta_{23}=0.7845$
\cite{Adamson:2011ch,Hosaka:2006zd}.  For the larger mass-squared difference, we set $\Delta_{32}=2.54 \times
10^{-3}$ eV$^2$, the value obtained by Daya Bay for normal hierarchy, so as to make our analysis  directly comparable to the Daya Bay analysis.

\begin{figure}[b]
\includegraphics*[width=3in]{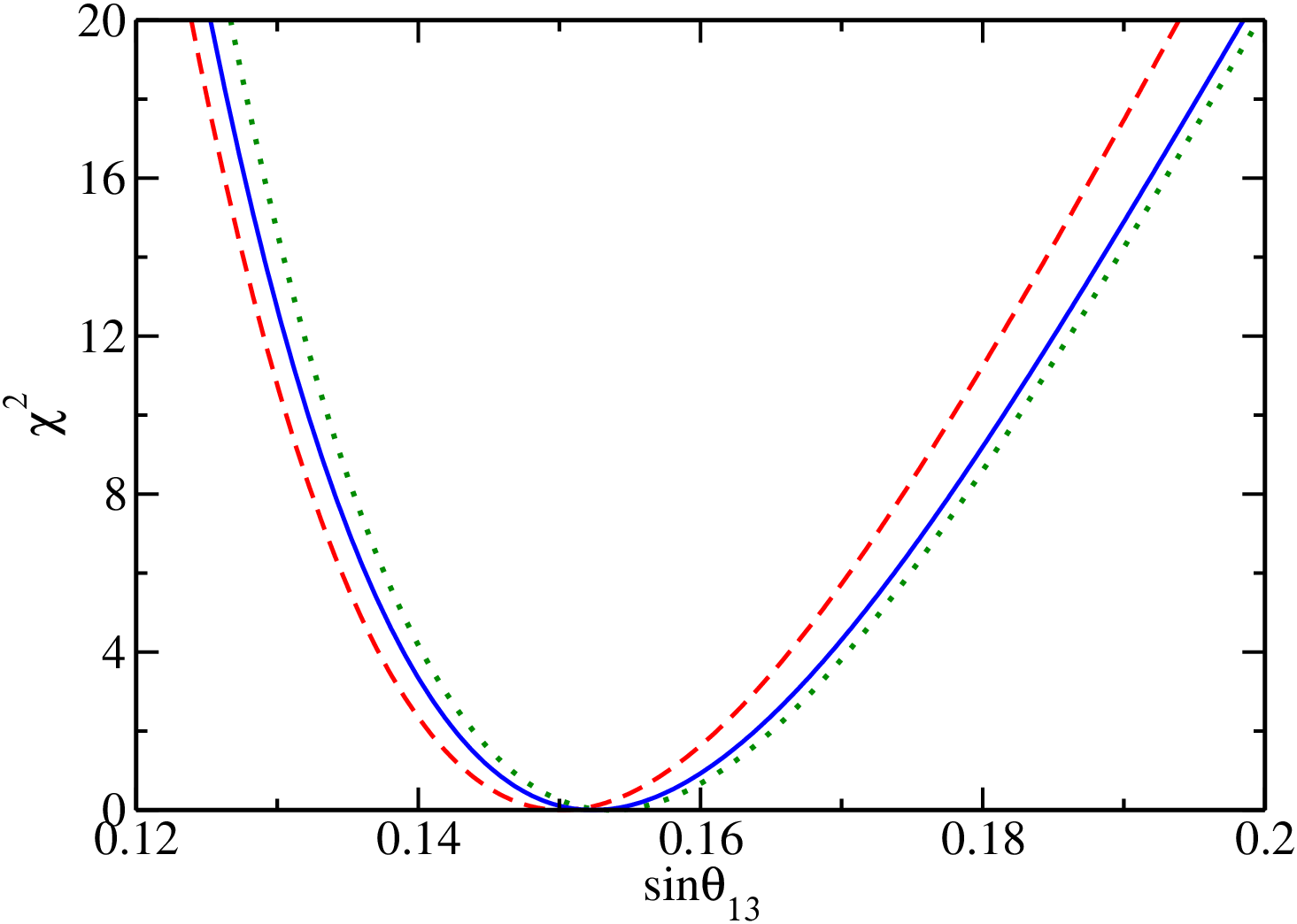}
\caption{$\Delta\chi^2$ versus $\sin\theta_{13}$ for the Daya Bay experiment. The solid (blue) curve is the result of an analysis
utilizing the exact oscillation probability, the dashed (red) curve uses the formula from \protect Eq.~(\ref{appee}), while the dotted (green) curve uses the exact oscillation probability, neglecting the $\Delta_{21}$ term.}
\label{fig2}
\end{figure}

In Fig.~\ref{fig2} we depict $\Delta\chi^2$ versus $\sin\theta_{13}$ for our analysis of the Daya Bay experiment. The
experimentalists find $\sin^2 2\,\theta_{13} = 0.090^{+0.009}_{-0.008}$ (one sigma errors) \cite{An:2013zwz}; we find the same result. The color code is
the same as for Fig.~\ref{fig1}.  The solid (blue) curve represents several indistinguishable results achieved by using: the exact expression, either hierarchy, with or without matter effects, or the use
of $\Delta_{ee}$ (with the $\phi_{21}$ oscillation term included).  The dashed (red) curve employs the approximation in Eq.~(\ref{appee}) and the dotted (green) curve employs the exact probability neglecting the $\phi_{21}$ oscillation term. 
The differences in the three $\Delta \chi^2$ curves might seem large for probabilities that
differ by tenths of a percent. But the parameter of interest, $\theta_{13}$, is not determined by ${\mathcal P}_{ee}$,
but rather by $\Delta\mathcal{P} =:1-{\mathcal P}_{ee}$. Since $\Delta\mathcal{P}$ peaks at around seven 
percent, the error for this quantity is larger in percent than the error in ${\mathcal P}_{ee}$ itself.  From this
result, we consider the use of the approximation from Ref.~\cite{Akhmedov:2004ny} or dropping the last term in Eq.~(\ref{exactp}) to
be inadequate.

\begin{figure}
\includegraphics*[width=3in]{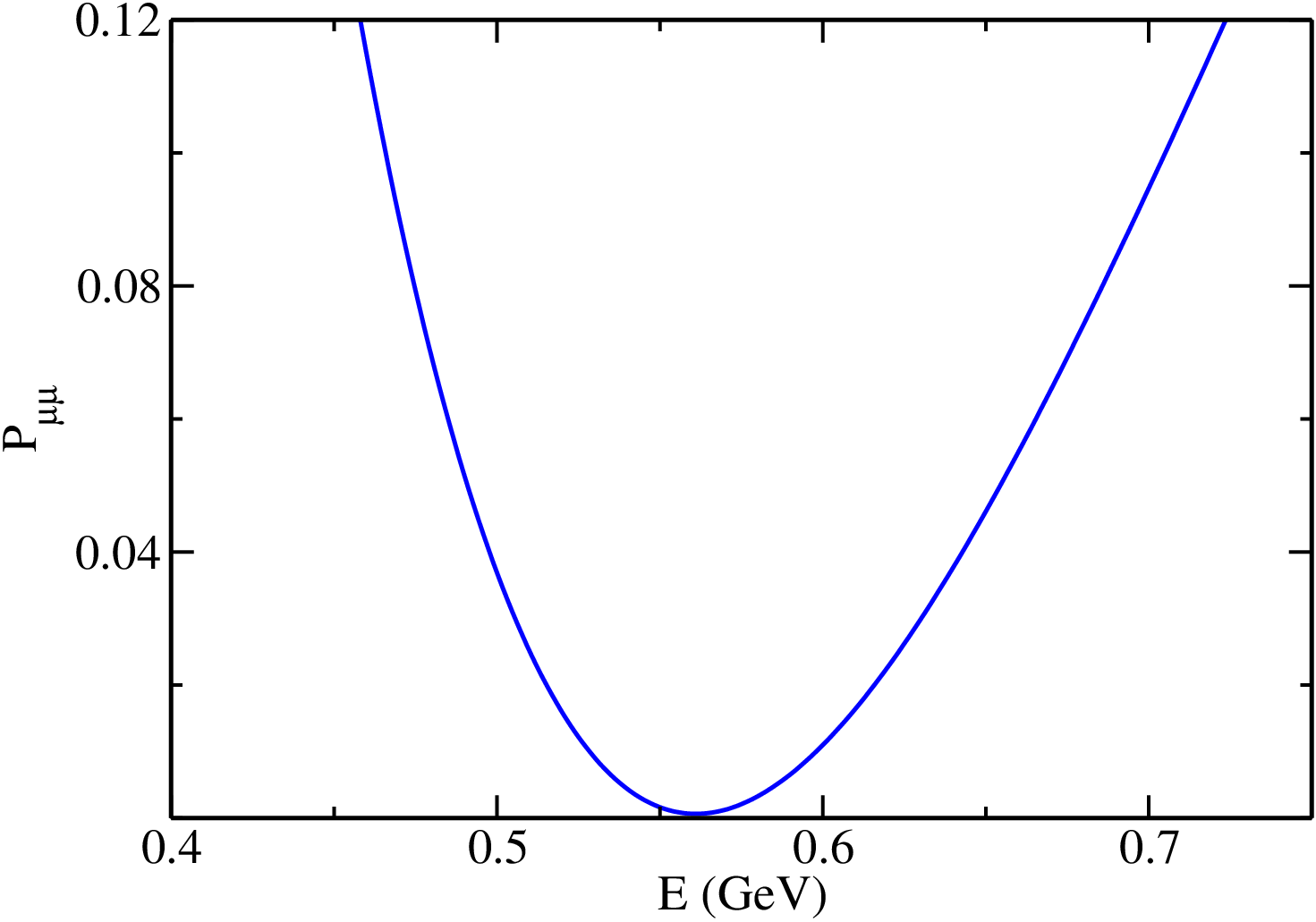}
\caption{The oscillation probability ${\mathcal P}_{\mu\mu}$ versus energy for energies near the oscillation minimum for
the T2K experiment with $L=295$ km. }
\label{fig3}
\end{figure}

\section{The T2K $\nu_\mu$ disappearance experiment and $\Delta_{32}$}

The long-baseline muon disappearance experiments,
MINOS \cite{Adamson:2014vgd} and T2K \cite{Abe:2013hdq}, and  the Super-K atmospheric experiment \cite{Hosaka:2006zd} yield clean measurements of the mass-squared difference $\Delta_{32}$. 
We will focus upon the T2K  experiment as it is poised to surpass MINOS' precision in the determination of $\Delta_{32}$ with its next data release.  In the T2K experiment, a beam of muon neutrinos travels over a 
long baseline through the Earth with an, assumed constant, density of 2.6 g/cm$^3$.    In Fig.~\ref{fig3} we depict the oscillation probability ${\mathcal P}_{\mu \mu}$ as a function of energy  in the region
near the minimum of the oscillation probability for the T2K experiment assuming a fixed baseline of 295 km. We focus on the minimum of the oscillation probability because we expect to find the differences between the exact and 
approximate oscillation formulae to be most significant and visible here. 
In the Figure, the curve represents the oscillation probability using the exact  numerical calculation for  vacuum oscillations, assuming normal hierarchy. The effects of matter or hierarchy 
are not sufficiently significant to be visible on this graph. For the inverse
hierarchy, there is a 10\% change in the probability at the minimum of $\mathcal{P}_{\mu\mu}$, corresponding to an
absolute change of only $6\times 10^{-5}$. At higher energies, there is a 0.4\% change. Matter results in a 20\% change at the minimum, corresponding to an absolute change of $1.5\times 10^{-4}$. At higher energies there is 
only a 0.02\% change.

From Ref.~\cite{Akhmedov:2004ny}, the approximate formula for the vacuum $\nu_\mu$ survival probability, ${\mathcal P}_{\mu\mu}$,  is:
\begin{eqnarray} {\mathcal P}_{\mu\mu} 
  &=& 1 - \sin^2 2\theta_{23}\,\sin^2\phi_{31}\nonumber\\
  &+&\alpha\, c^2_{12}\,\sin^2 2\theta_{23}\,\,\phi_{31}\,\sin 2\phi_{31}\nonumber\\
  &-&\alpha^2\,c^2_{23}\sin^2 2\theta_{12}\,\phi^2_{31}\nonumber\\
  &-&\alpha^2\,c^2_{12}\,\sin^2 2 \theta_{23}\,\phi^2_{31}\,(\cos2\phi_{31}-s_{12}^2)\nonumber\\
  &-&4\,s^2_{13}\,s^2_{23}\,\sin^2\phi_{31}\nonumber\\
  &+&2\, s^2_{13}\,\sin^22\theta_{23}\,\sin^2\phi_{31}\nonumber\\
  &-&2\,\alpha\,s_{13}\,\sin2\theta_{12}\,\sin2\theta_{23}\,s_{23}^2\,\phi_{31}\,\sin 2\phi_{31}.
  \label{appmm}
\end{eqnarray}
This is a very good approximation to use for the T2K experiment. Differences between the exact and approximate oscillation probabilities are so small that they are not visible on the graph in Fig.~\ref{fig3}.  At the
minimum of $\mathcal{P}_{\mu\mu}$ the change is 10\%, with an absolute change of $6\times 10^{-5}$. At higher energies, the change is 0.4\%.

\begin{figure}			
\includegraphics*[width=3in]{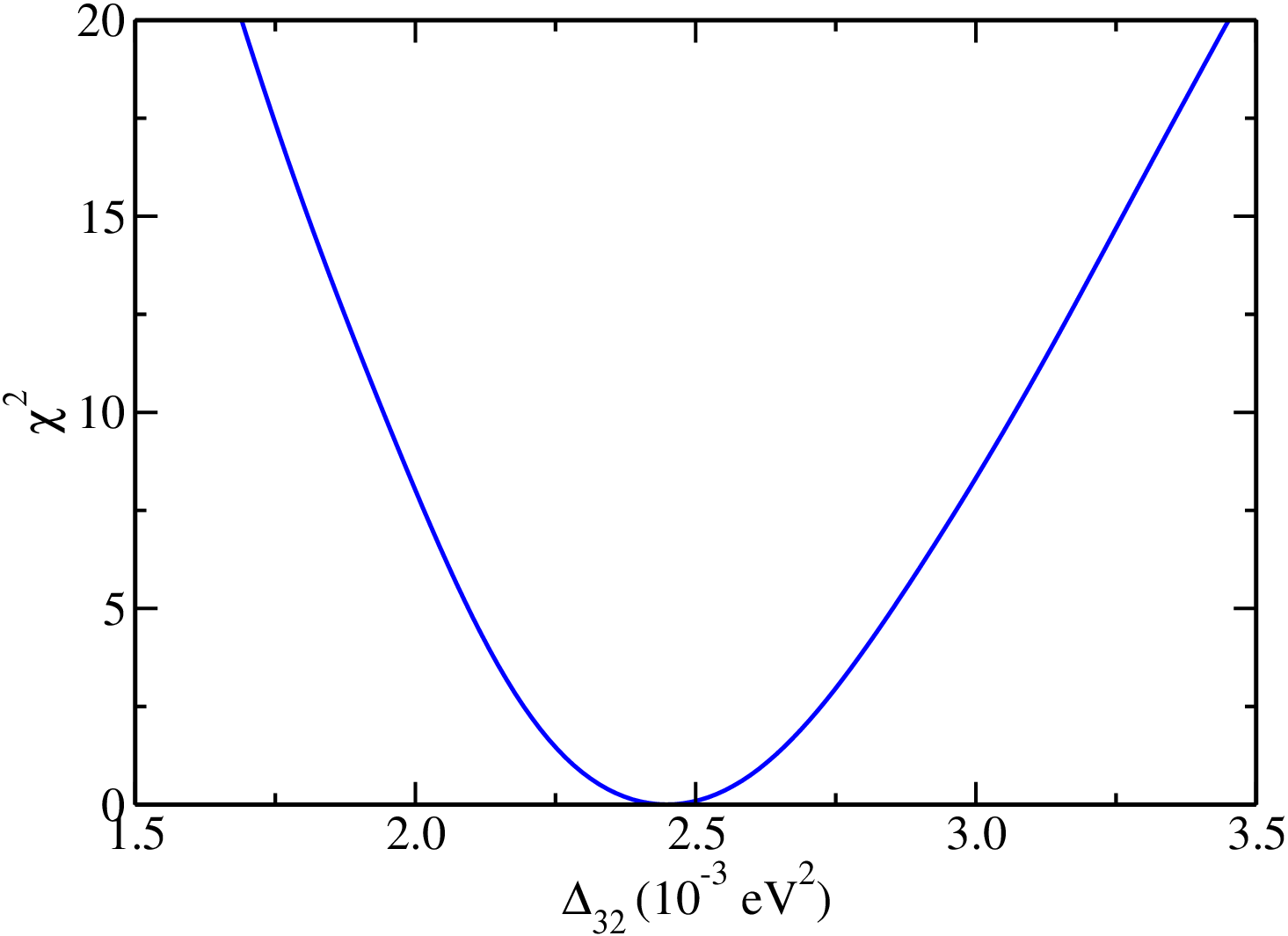}
\caption{$\Delta\chi^2$ versus neutrino mass-squared difference $\Delta_{32}$ for the T2K disappearance experiment.
The curve represents the results utilizing the exact probability, either hierarchy, with and without matter effects, and
also the use of the approximate oscillation probability given in 
\protect Eq.~(\ref{appmm}). }
\label{fig4}
\end{figure}

For the T2K \cite{Abe:2013hdq} disappearance experiment, $\Delta \chi^2$ versus $\Delta_{32}$ is shown in Fig.~\ref{fig4}. We find that all oscillation probabilities yield equivalent results.
That is, the  curve represents the exact results of our model for normal and inverse hierarchy, either in vacuum or in constant density matter. The curve also
represents the results of using the approximate oscillation formula from Ref.~\cite{Akhmedov:2004ny}. 
We have checked to see if any of the terms in this formula might be neglected and found that all are needed to maintain an accuracy better
than 0.4\%, i.e., less than about a half line width on the plots. There are terms which are zero
for $\theta_{23}$ set to maximal mixing, but these cannot be neglected if one wants a generally applicable code. 

\section{The T2K $\nu_e$ appearance experiment and hierarchy}

The T2K disappearance and Daya Bay experiments are predominantly sensitive to a single parameter, $\Delta_{32}$ and $\theta_{13}$, but with 
sensitivity to a second parameter, $\theta_{23}$ \cite{Abe:2014ugx} and $\Delta_{32}$, respectively.
The $\nu_\mu \to \nu_e$ appearance probability, ${\mathcal P}_{\mu e}$, measured in the T2K appearance experiment  is of a different
character \cite{Abe:2014ugx}. ${\mathcal P}_{\mu
e} $ is sensitive to matter effects, to the hierarchy of the neutrino oscillations, and to the Dirac CP phase. In order to be able to provide information about these quantities, input is needed from other neutrino 
oscillation experiments. In what follows, we will assume there is no CP violation and examine the question of hierarchy in the context of the T2K appearance experiment.

\begin{figure}
\includegraphics*[width=3in]{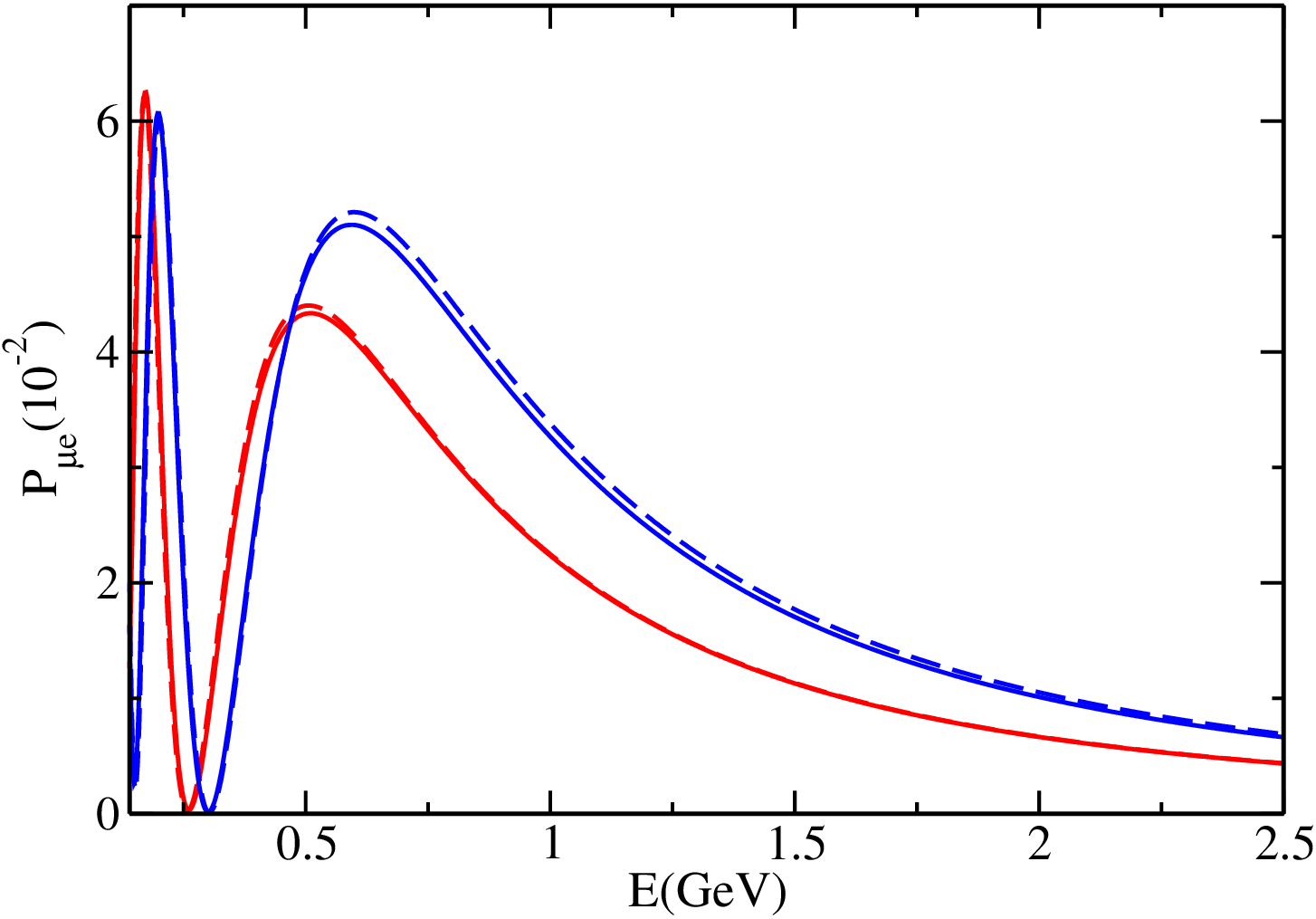}
\caption{The oscillation probability ${\mathcal P}_{\mu e}$ as a function of energy over the energy range appropriate
for the T2K appearance experiment. The blue curves employ the normal hierarchy, and red employ the inverse hierarchy.  Solid curves use the
exact oscillation probability and dashed curves  use the approximate probability given in Eq.~(\ref{appme}).}
\label{fig5}
\end{figure}

As stated earlier, hierarchy refers to the ordering of the mass eigenstates.  In particular, is $m_3$ less than or greater than $m_1,m_2$?   In principle, this can be determined from the T2K appearance data.  As shown in 
Fig.~\ref{fig5}, ${\mathcal P}_{\mu e}$ is small, only a few percent, in the region where we can presently measure it.
This makes getting good statistics difficult.  For T2K
the detector is located off-axis in order to reduce the background, which also reduces the overall flux, while MINOS \cite{Adamson:2013ue}, is on-axis producing a
background that is comparable to the signal. 
The recent T2K results are quite exciting as they are the first
measurements that provide clues to the hierarchy question and the existence of CP violation. This is only true because of the relatively large value of
$\theta_{13}$ discovered by Daya Bay \cite{An:2013zwz} and RENO \cite{Ahn:2012nd}. 

The approximate oscillation probability for ${\mathcal P}_{\mu e}$ from Ref.~\cite{Akhmedov:2004ny} is given by
\begin{eqnarray}
{\mathcal P}_{\mu e} &=& \alpha^2\,c^2_{23}\,\sin^2 2\theta_{12}\,\frac{\sin^2 (A\, \phi_{31})}{A^2} \nonumber\\
                     &+& 4\,s^2_{13}\,s^2_{23}\, \frac{\sin^2((A-1)\phi_{31})}{(A-1)^2}\nonumber\\
		     &+& 2\,\alpha\,s_{13}\,\sin2\theta_{12}\,\sin2\theta_{23}\cos\phi_{31}\,\nonumber\\
		     &\times&\,\sin(A\phi_{31} )\,\frac{\sin((A-1)\phi_{31})}{A(A-1)}\,\,
\label{appme}
\end{eqnarray}
where matter effects are included in the factor $A:= 2\,E\,V/\Delta_{31}$ with  $V$ the MSW potential in eV.

In Fig.~\ref{fig5} we depict the oscillation probability ${\mathcal P}_{\mu e}$ versus energy for the energy range
relevant to T2K. The solid curves are exact results, dashed curves the approximation given in
Eq.~(\ref{appme}). The blue curves are normal hierarchy, the red inverse hierarchy.

\begin{figure}
\includegraphics*[width=3in]{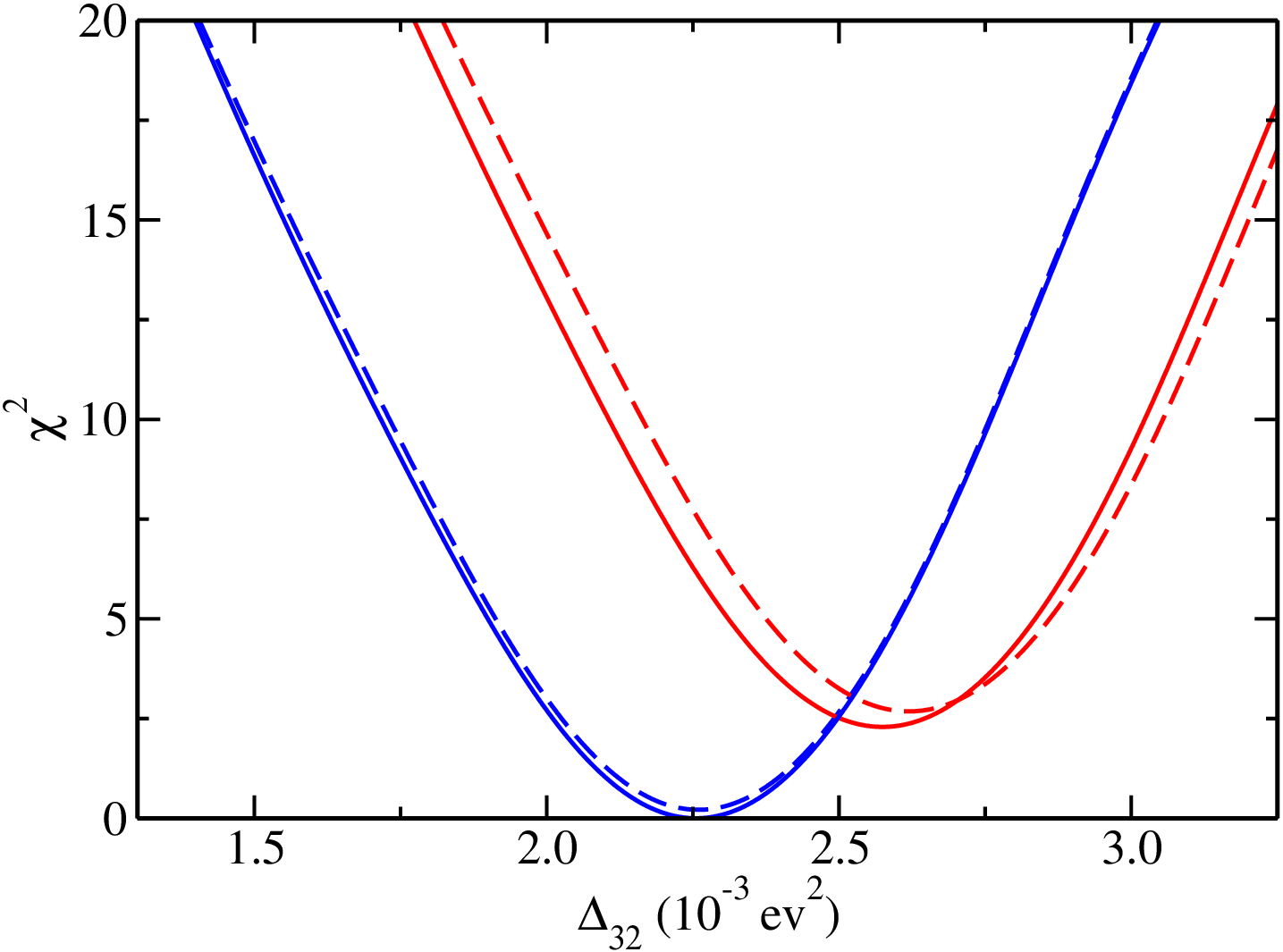}
\caption{$\Delta\chi^2$ versus neutrino mass-squared difference $\Delta_{32}$ for the T2K appearance experiment.
The solid curves represent the results of utilizing the exact probability; the dashed curves represent results from
utilizing the approximate probability from Eq.~(\ref{appme}). The blue curves employ normal hierarchy; the red
curves employ inverse hierarchy. }
\label{fig6}
\end{figure}

We see the significant hierarchy dependence. The approximate curves are close to the exact curves. For the normal
hierarchy, the approximate curve differs from the exact by 3\% near the peak, and 4\% at higher energies. For the
inverse hierarchy, the difference is 2\% near the peak and 4\% at the higher energies.

In Fig.~\ref{fig6} we show $\Delta\chi^2$ versus $\Delta_{32}$ for our analysis of the T2K appearance experiment. 
The solid curves represent the results utilizing the exact probability; the dashed curves represent results from
utilizing the approximate probability from Eq.~(\ref{appme}); the blue curves are normal hierarchy; the red
curves are inverse hierarchy. For normal hierarchy, the exact curve gives $\Delta_{32} = 2.26\pm 0.26\times 10^{-3}$
eV$^2$ and the approximate curve gives $2.26\pm 0.25\times 10^{-3}$ eV$^2$, a surprisingly good agreement given the
difference in the oscillation probabilities. However, for the inverse
hierarchy, the exact curve gives $\Delta_{32} = 2.57\pm 0.26\times 10^{-3}$ eV$^2$ and the approximate curve gives
$2.62\pm 0.26\times 10^{-3}$ eV$^2$. For the inverse hierarchy, there is a shift in the best fit value for $\Delta_{32}$
of $0.05\times 10^{-3}$ eV$^2$ and a shift in the value of $\chi^2$ at the minimum of 0.4. This is not large in an absolute
sense, but if one wishes to maintain a limit of one percent error on the actual theoretical calculations, this is 
not acceptable. 

\section{Discussion}

We have examined the use of approximate oscillation probabilities for ${\mathcal P}_{ee}$ in the
context of analyzing the Daya Bay \cite{Akhmedov:2004ny} experiment, ${\mathcal P}_{\mu\mu}$ in the context of analyzing
the T2K disappearance experiment \cite{Abe:2013hdq}, and ${\mathcal P}_{\mu e}$ in the context of analyzing the T2K appearance
experiment \cite{Abe:2013hdq}. Since matter effects for Daya Bay are quite small, the rather simple exact formula for ${\mathcal
P}_{ee}$ in vacuum can be used. The approximation used by Daya Bay that uses only one mass-squared difference
$\Delta_{ee}$ is accurate as long as the $\phi_{21}$ mixing term is included. For the T2K disappearance experiments, 
the approximation formula of Ref.~\cite{Akhmedov:2004ny} is
found to be quite accurate, with the vacuum form given in Eq.~(\ref{appmm}) adequate. All terms in the rather lengthy
formula must be included. For the T2K appearance experiment, matter effects must be included. The formula,
Eq.~(\ref{appme}), is found to give reasonable results for the normal hierarchy but not so for the inverse hierarchy.  Based on these results, we find that our investigation raises a number of questions.

First, are these formulae adequate for learning about CP violation? The recent T2K
results \cite{Abe:2013hdq} are sensitive to the CP phase and, hence, when combined with reactor data is able to rule out some values of the phase $\delta_{CP}$.  Since we are not satisfied by the 
ability of the approximate formula to discern between hierarchies and since the hierarchy question
is deciding between two distinct answers, the approximate formula is certainly not satisfactory to measure the value of
the continuous variable $\delta_{CP}$.

The second question is whether or not approximate oscillation probabilities are adequate for the analysis of atmospheric
data \cite{Hosaka:2006zd}. Atmospheric data are sensitive to both the hierarchy and the CP
phase. Can we reliably analyze atmospheric data well enough such that we can
extract this information? There exist two published analyses \cite{Capozzi:2013csa, GonzalezGarcia:2012sz} which yield
qualitatively different results when addressing the questions of hierarchy and the CP phase. The difference in these results may be due to the use of different approximations for
the oscillation probabilities. As it stands, it seems that the approximate probabilities are not adequate for atmospheric data. On the other hand, the atmospheric data has become less significant 
with the recent, more accurate data from Daya Bay
and T2K. The atmospheric data does still affect the final values of $\Delta_{32}$ and $\theta_{23}$, but data coming in
from T2K and future data from NO$\nu$A \cite{Jediny:2014lda} will render the contribution from atmospheric data negligible, as Daya Bay has
done for $\theta_{13}$. 

Third, what is the impact of using various approximate formulae on the consistency of neutrino oscillation analyses? Two points are worth noting.  First, different approximations used by different experimental 
groups applied to data used to constrain the same oscillation parameter, such as the reactor constraints on $\theta_{13}$, may lead to inconsistencies, especially as data becomes more precise.  A second, and far more 
subtle, point is that changing the approximation applied to an updated data set by the same experimental group merits careful interpretation and comparison to previous analyses.  The only statistically robust, reproducible 
analysis is a full exact three neutrino calculation using a numerical code for the case where matter effects are included, or a full exact formula for the case of vacuum.  It is not immediately obvious whether or not old and 
new data releases may appear falsely consistent or inconsistent partly as a result of changing approximations, especially as systematic errors decrease over time.  Furthermore, this question reiterates the point discussed 
above that a possible source of the discrepancy between global analyses is the use of different approximations, particularly for the atmospheric data, which covers a broad range of $L/E$.  This leads to one final issue.

For data sets sensitive to many types of oscillation physics, such 
as matter and hierarchy effects, it is difficult to find or construct reliable approximations that take into account multiple sub-leading effects simultaneously.  The danger in using an overly-tailored approximation, designed 
to constrain only one specific parameter, is the loss of correlations between sub-leading effects which can enrich the final result, much like combining complimentary data sets is more powerful than relying on one data set alone.  
For data sets which are rich in subtle physics, such as the long-baseline and atmospheric data that are sensitive to matter effects, the CP phase, and the mass ordering, the loss of information resulting from the use of 
approximate oscillation probabilities has yet to be sufficiently assessed or quantified and, hence, merits continued, careful examination.

\section{Conclusions}

As neutrino oscillation experiments become increasingly accurate, the need to ensure the robustness and accuracy of analyses of the data becomes more important. Our examination of the variation between analysis outcomes of the 
same data using different approximate oscillation formulae indicates that the impact of the oscillation formula used on the final results of an analysis should not be underestimated. We stress that our study indicates that the 
implicit definition of the term ``accuracy of an approximate probability" should be expanded to include not only how much it varies from the exact probability but also how robust a statistical analysis of the data is compared to 
one conducted with the exact probability. The need for continued, thorough pre-assessment of the validity and reliability of approximate oscillation formulae will become increasingly mandatory as systematic errors reach percent 
level and the signatures of ever more subtle neutrino physics is sought. Already, small discrepancies among global analyses suggest that the data has become precise enough that the ``sub-leading effects" picture, which implied 
the permissible use of expansions that neglect small terms, has outlived its usefulness. For this reason we encourage a shift from the term ``sub-leading effects" to ``subtle physics," which implies the need for a more careful 
and complete treatment of the data.

Fortunately, transitioning from the past era of approximations to the  new era of using exact oscillation formulae is simple. To contribute to this effort we will publish the short, concise subroutine implemented here that uses 
the method of Ref.~\cite{Ohlsson:1999xb} for calculating oscillation
probabilities, including constant matter density and CP effects. The subroutine is stable, efficient, and accurate and thus removes the need to use approximations for the analysis of long-baseline, atmospheric, and solar data. 
Furthermore, exact vacuum expressions for electron antineutrino disappearance are trivially short and, hence, negate the need to use approximate expressions for analyses of reactor data. The benefit is that progressing to the 
use of exact oscillation formulae, which is not a computationally burdensome change, will allow future published analyses of neutrino oscillation data to be more robust, more consistent, and more sensitive to the subtle physics 
of the neutrino mass ordering and CP phase that precision measurement are beginning to make accessible.

\bibliography{approx.bib}

\begin{thebibliography}{16}
\expandafter\ifx\csname natexlab\endcsname\relax\def\natexlab#1{#1}\fi
\expandafter\ifx\csname bibnamefont\endcsname\relax
  \def\bibnamefont#1{#1}\fi
\expandafter\ifx\csname bibfnamefont\endcsname\relax
  \def\bibfnamefont#1{#1}\fi
\expandafter\ifx\csname citenamefont\endcsname\relax
  \def\citenamefont#1{#1}\fi
\expandafter\ifx\csname url\endcsname\relax
  \def\url#1{\texttt{#1}}\fi
\expandafter\ifx\csname urlprefix\endcsname\relax\def\urlprefix{URL }\fi
\providecommand{\bibinfo}[2]{#2}
\providecommand{\eprint}[2][]{\url{#2}}

\bibitem[{\citenamefont{An et~al.}(2014)}]{An:2013zwz}
\bibinfo{author}{\bibfnamefont{F.}~\bibnamefont{An}} \bibnamefont{et~al.}
  (\bibinfo{collaboration}{Daya Bay Collaboration}),
  \bibinfo{journal}{Phys.~Rev.~Lett.~} \textbf{\bibinfo{volume}{112}},
  \bibinfo{pages}{061801} (\bibinfo{year}{2014}).

\bibitem[{\citenamefont{Adamson et~al.}()}]{Adamson:2014vgd}
\bibinfo{author}{\bibfnamefont{P.}~\bibnamefont{Adamson}} \bibnamefont{et~al.}
  (\bibinfo{collaboration}{MINOS Collaboration}),
  \bibinfo{note}{arXiv:1403.0867 (2014)}.

\bibitem[{\citenamefont{Abe et~al.}(2013)}]{Abe:2013fuq}
\bibinfo{author}{\bibfnamefont{K.}~\bibnamefont{Abe}} \bibnamefont{et~al.}
  (\bibinfo{collaboration}{T2K Collaboration}),
  \bibinfo{journal}{Phys.~Rev.~Lett.~} \textbf{\bibinfo{volume}{111}},
  \bibinfo{pages}{211803} (\bibinfo{year}{2013}).

\bibitem[{\citenamefont{Jediný}(2014)}]{Jediny:2014lda}
\bibinfo{author}{\bibfnamefont{F.}~\bibnamefont{Jediný}}
  (\bibinfo{collaboration}{NOvA}), \bibinfo{journal}{J.~Phys.~Conf.~Ser.~}
  \textbf{\bibinfo{volume}{490}}, \bibinfo{pages}{012019}
  (\bibinfo{year}{2014}).

\bibitem[{\citenamefont{Akhmedov et~al.}(2004)\citenamefont{Akhmedov,
  Johansson, Lindner, Ohlsson, and Schwetz}}]{Akhmedov:2004ny}
\bibinfo{author}{\bibfnamefont{E.~K.} \bibnamefont{Akhmedov}},
  \bibinfo{author}{\bibfnamefont{R.}~\bibnamefont{Johansson}},
  \bibinfo{author}{\bibfnamefont{M.}~\bibnamefont{Lindner}},
  \bibinfo{author}{\bibfnamefont{T.}~\bibnamefont{Ohlsson}}, \bibnamefont{and}
  \bibinfo{author}{\bibfnamefont{T.}~\bibnamefont{Schwetz}},
  \bibinfo{journal}{JHEP} \textbf{\bibinfo{volume}{0404}}, \bibinfo{pages}{078}
  (\bibinfo{year}{2004}).

\bibitem[{\citenamefont{Ohlsson and Snellman}(2000)}]{Ohlsson:1999xb}
\bibinfo{author}{\bibfnamefont{T.}~\bibnamefont{Ohlsson}} \bibnamefont{and}
  \bibinfo{author}{\bibfnamefont{H.}~\bibnamefont{Snellman}},
  \bibinfo{journal}{J.~Math.~Phys.~} \textbf{\bibinfo{volume}{41}},
  \bibinfo{pages}{2768} (\bibinfo{year}{2000}).

\bibitem[{\citenamefont{Abe et~al.}(2014{\natexlab{a}})}]{Abe:2013hdq}
\bibinfo{author}{\bibfnamefont{K.}~\bibnamefont{Abe}} \bibnamefont{et~al.}
  (\bibinfo{collaboration}{T2K Collaboration}),
  \bibinfo{journal}{Phys.~Rev.~Lett.~} \textbf{\bibinfo{volume}{112}},
  \bibinfo{pages}{061802} (\bibinfo{year}{2014}{\natexlab{a}}).

\bibitem[{\citenamefont{Minakata et~al.}(2006)\citenamefont{Minakata, Nunokawa,
  Parke, and Zukanovich~Funchal}}]{Minakata:2006gq}
\bibinfo{author}{\bibfnamefont{H.}~\bibnamefont{Minakata}},
  \bibinfo{author}{\bibfnamefont{H.}~\bibnamefont{Nunokawa}},
  \bibinfo{author}{\bibfnamefont{S.~J.} \bibnamefont{Parke}}, \bibnamefont{and}
  \bibinfo{author}{\bibfnamefont{R.}~\bibnamefont{Zukanovich~Funchal}},
  \bibinfo{journal}{Phys.~Rev.~} \textbf{\bibinfo{volume}{D74}},
  \bibinfo{pages}{053008} (\bibinfo{year}{2006}).

\bibitem[{\citenamefont{Beringer et~al.}(2012)}]{Beringer:1900zz}
\bibinfo{author}{\bibfnamefont{J.}~\bibnamefont{Beringer}} \bibnamefont{et~al.}
  (\bibinfo{collaboration}{Particle Data Group}),
  \bibinfo{journal}{Phys.~Rev.~} \textbf{\bibinfo{volume}{D86}},
  \bibinfo{pages}{010001} (\bibinfo{year}{2012}).

\bibitem[{\citenamefont{Adamson et~al.}(2011)}]{Adamson:2011ch}
\bibinfo{author}{\bibfnamefont{P.}~\bibnamefont{Adamson}} \bibnamefont{et~al.}
  (\bibinfo{collaboration}{MINOS Collaboration}), \bibinfo{journal}{Phys.Rev.}
  \textbf{\bibinfo{volume}{D84}}, \bibinfo{pages}{071103}
  (\bibinfo{year}{2011}).

\bibitem[{\citenamefont{Hosaka et~al.}(2006)}]{Hosaka:2006zd}
\bibinfo{author}{\bibfnamefont{J.}~\bibnamefont{Hosaka}} \bibnamefont{et~al.}
  (\bibinfo{collaboration}{Super-Kamiokande Collaboration}),
  \bibinfo{journal}{Phys.~Rev.~} \textbf{\bibinfo{volume}{D74}},
  \bibinfo{pages}{032002} (\bibinfo{year}{2006}).

\bibitem[{\citenamefont{Abe et~al.}(2014{\natexlab{b}})}]{Abe:2014ugx}
\bibinfo{author}{\bibfnamefont{K.}~\bibnamefont{Abe}} \bibnamefont{et~al.}
  (\bibinfo{collaboration}{T2K Collaboration})
  (\bibinfo{year}{2014}{\natexlab{b}}), \eprint{1403.1532}.

\bibitem[{\citenamefont{Adamson et~al.}(2013)}]{Adamson:2013ue}
\bibinfo{author}{\bibfnamefont{P.}~\bibnamefont{Adamson}} \bibnamefont{et~al.}
  (\bibinfo{collaboration}{MINOS Collaboration}),
  \bibinfo{journal}{Phys.~Rev.~Lett.~} \textbf{\bibinfo{volume}{110}},
  \bibinfo{pages}{171801} (\bibinfo{year}{2013}).

\bibitem[{\citenamefont{Ahn et~al.}(2012)}]{Ahn:2012nd}
\bibinfo{author}{\bibfnamefont{J.}~\bibnamefont{Ahn}} \bibnamefont{et~al.}
  (\bibinfo{collaboration}{RENO collaboration}),
  \bibinfo{journal}{Phys.~Rev.~Lett.~} \textbf{\bibinfo{volume}{108}},
  \bibinfo{pages}{191802} (\bibinfo{year}{2012}).

\bibitem[{\citenamefont{Capozzi et~al.}(2013)\citenamefont{Capozzi, Fogli,
  Lisi, Marrone, Montanino et~al.}}]{Capozzi:2013csa}
\bibinfo{author}{\bibfnamefont{F.}~\bibnamefont{Capozzi}},
  \bibinfo{author}{\bibfnamefont{G.}~\bibnamefont{Fogli}},
  \bibinfo{author}{\bibfnamefont{E.}~\bibnamefont{Lisi}},
  \bibinfo{author}{\bibfnamefont{A.}~\bibnamefont{Marrone}},
  \bibinfo{author}{\bibfnamefont{D.}~\bibnamefont{Montanino}},
  \bibnamefont{et~al.} (\bibinfo{year}{2013}), \bibinfo{note}{arXiv:1312.2878}.

\bibitem[{\citenamefont{Gonzalez-Garcia
  et~al.}(2012)\citenamefont{Gonzalez-Garcia, Maltoni, Salvado, and
  Schwetz}}]{GonzalezGarcia:2012sz}
\bibinfo{author}{\bibfnamefont{M.}~\bibnamefont{Gonzalez-Garcia}},
  \bibinfo{author}{\bibfnamefont{M.}~\bibnamefont{Maltoni}},
  \bibinfo{author}{\bibfnamefont{J.}~\bibnamefont{Salvado}}, \bibnamefont{and}
  \bibinfo{author}{\bibfnamefont{T.}~\bibnamefont{Schwetz}},
  \bibinfo{journal}{JHEP} \textbf{\bibinfo{volume}{1212}}, \bibinfo{pages}{123}
  (\bibinfo{year}{2012}).

\end{thebibliography}

\end{document}